\documentclass[sigconf,screen,urlbreakonhyphens]{acmart}

\copyrightyear{2026}
\acmYear{2026}
\setcopyright{cc}
\setcctype{by}
\acmConference[DAC '26]{63rd ACM/IEEE Design Automation Conference}{July 26--29, 2026}{Long Beach, CA, USA}
\acmBooktitle{63rd ACM/IEEE Design Automation Conference (DAC '26), July 26--29, 2026, Long Beach, CA, USA}
\acmDOI{10.1145/3770743.3803967}
\acmISBN{979-8-4007-2254-7/2026/07}

\usepackage{tikz}
\newcommand{\fullcirc}{\raisebox{0.7ex}{\tikz[baseline=(char.base)]{\node[shape=circle, fill=black, inner sep=1.5pt] (char) {};}}}
\newcommand{\halfcirc}{\raisebox{0.7ex}{\tikz[baseline=(char.base)]{\node[shape=circle, draw=black, inner sep=1.5pt] (char) {}; \node[shape=circle, fill=black, inner sep=0.8pt] at (char.center) {};}}}
\newcommand{\emptycirc}{\raisebox{0.7ex}{\tikz[baseline=(char.base)]{\node[shape=circle, draw=black, inner sep=1.5pt] (char) {};}}}

\usepackage{siunitx}

\begin{document}
	
\title[Toki: Profiling HBM Performance on FPGA Systems with RISC-V Soft Cores and PCIe Host DMA Traffic]{Toki: Profiling HBM Performance on FPGA Systems with RISC-V Soft Cores and PCIe Host DMA Traffic}

\author{Andrea Galimberti}
\orcid{0000-0003-0254-3933}
\affiliation{%
	\institution{Politecnico di Milano}
	\city{Milan}
	\country{Italy}
}
\email{andrea.galimberti@polimi.it}

\author{Andrea Motta}
\orcid{0009-0008-2300-2785}
\affiliation{%
	\institution{Politecnico di Milano}
	\city{Milan}
	\country{Italy}
}
\email{andrea.motta@polimi.it}

\author{Gianni Antichi}
\orcid{0000-0002-6063-4975}
\affiliation{%
	\institution{Politecnico di Milano}
	\city{Milan}
	\country{Italy}
}
\email{gianni.antichi@polimi.it}

\author{Davide Zoni}
\orcid{0000-0002-9951-062X}
\affiliation{%
	\institution{Politecnico di Milano}
	\city{Milan}
	\country{Italy}
}
\email{davide.zoni@polimi.it}

\begin{abstract}
Programmable RISC-V soft cores are becoming more widespread in data-center scenarios, making it crucial to design efficient systems that deploy them on FPGA chips with HBM memory.
Toki, released as open source, is the first hardware-software framework that enables profiling the performance of HBM on FPGA accelerator cards by jointly considering
(i) the execution of workloads on RISC-V soft cores instantiated on the FPGA and
(ii) the injection of memory traffic from the host system via DMA over PCIe,
providing insights that cannot be obtained with synthetic traffic generators alone.
Extensive experiments target an AMD Alveo U55C card, deploying up to 60 RISC-V compute cores and stressing its HBM2 memory through real-world applications and microbenchmarks with user-defined access patterns.
Results showcase how Toki can effectively profile the impact on HBM performance of the compute cores' organization, the workload's memory access patterns, data locality, contention over the memory controllers, and host traffic.%
\end{abstract}

\ccsdesc[500]{Hardware~Reconfigurable logic and FPGAs}
\ccsdesc[500]{Hardware~Dynamic memory}
\ccsdesc[300]{Hardware~Buses and high-speed links}
\ccsdesc[300]{Computer systems organization~Multicore architectures}

\keywords{FPGA, HBM, RISC-V, Performance profiling, Memory systems, Soft cores, PCI Express, Direct memory access}

\received{18 November 2025}
\received[accepted]{23 February 2026}

\maketitle

\section{Introduction}
\label{sec:introduction}
Modern data centers are tasked with managing increasingly demanding workloads, such as AI, real-time data analytics, and HPC, that require both high computational power and efficient memory systems to ensure scalability, energy efficiency, and quality of service. FPGAs have emerged as essential components in this context due to their flexibility~\cite{Bobda_2022CSUR}, enabling the deployment of custom accelerators that exploit massive parallelism and can be dynamically reprogrammed to adapt to evolving application requirements~\cite{Montanaro_2025TC}.

While computational throughput has increased rapidly, memory bandwidth has not scaled at the same rate, creating a critical bottleneck for data-intensive applications. HBM addresses this challenge by using 3D packaging to stack memory dies vertically and connect them to processing logic through a silicon interposer with wide I/O interfaces, achieving industry-leading data throughput while maintaining moderate latency~\cite{Kim_2016MSSC}.
Prior research has explored a wide range of HBM-enabled FPGA computing solutions. Application-specific architectures include accelerators for graph processing~\cite{Hu_2021ICCAD} and breadth-first search in graphs~\cite{Li_2024TRETS}, as well as overlay processors for graph neural networks~\cite{Tang_2024TRETS}. Other efforts have used HBM for accelerating sparse matrix-vector multiplication~\cite{Jain_2023FCCM,Rajashekar_2024FPGA,Tareen_2024FCCM,Yi_2024DATE,Sonmez_2025FCCM}, matrix transposition~\cite{Yang_2025FCCM}, LLM inference~\cite{Zeng_2024FPGA}, fully homomorphic encryption~\cite{Xu_2025FPGA}, and wave simulations~\cite{Gourounas_2025FCCM}.

\begin{figure}[t]
	\centering
	\includegraphics[width=0.94\columnwidth]{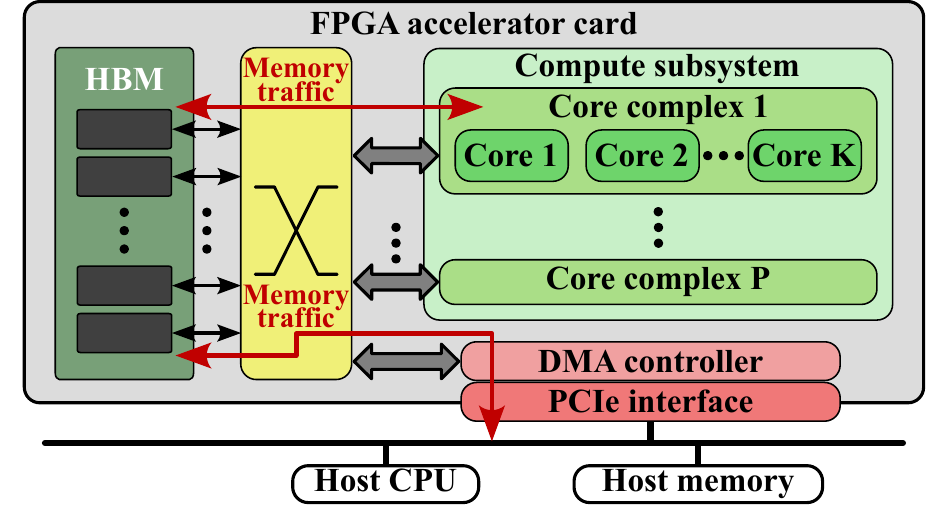}
	\caption{Toki's high-level architecture. Toki profiles HBM performance by concurrently stressing it with traffic from
		RISC-V soft cores and from the PCIe host system.}
	\label{fig:arch_top}
\end{figure}

Emerging trends have demonstrated the viability of using soft-core RISC-V CPUs instantiated on FPGAs as the primary compute elements for systems where programmability, flexibility, and rapid development are prioritized over raw peak performance~\cite{Ahn_2025FCCM}. This design approach offers a unified software-programmable architecture that simplifies development and supports rapid iteration across a broad range of applications. When deployed alongside HBM, these systems can exploit high memory parallelism while maintaining flexibility at the software level~\cite{Iskandar_2022TRETS}. As a result, several recent architectures have explored different strategies for scaling RISC-V cores on HBM-equipped FPGAs.
Notable examples include Vortex~\cite{Ahn_2025FCCM}, a soft-core GPU with support for CUDA kernels; Phalanx~\cite{Gray_2016FCCM,Gray_2019H2RC}, a kilo-core overlay processor; SPARKLE~\cite{Abdelhamid_2024MCSoC-SPARKLE}, a 1024-core architecture operating at 400\,MHz that leverages barrel processing; and the 32-core overlay processor in~\cite{Elshimy_2023MCSoC}, where each pair of cores has a dedicated memory controller to maximize HBM bandwidth utilization.

Despite the diversity and scale of these architectures built around soft-core CPUs on FPGAs equipped with HBM, prior efforts have largely focused on design innovation, whereas no systematic framework or platform currently exists to profile the performance of such systems. Existing solutions focus on other memory technologies such as DDR4~\cite{Galimberti_2025ISCAS}, isolated aspects of memory performance~\cite{Olgun_2023TCAD}, or accelerator workloads~\cite{Huang_2022TC}, but they all fail to capture the complex interactions between HBM, soft-core CPUs deployed in FPGA logic, and host systems.
Conversely, with the growing adoption of RISC-V architectures on FPGA-based platforms, understanding how these architectures perform in conjunction with HBM and PCIe host traffic is critical for optimizing system-level performance.

\subsection*{Contributions}
This paper introduces Toki, the first hardware-software (HW/SW) framework designed to profile the performance of HBM on FPGA accelerator cards by jointly considering two critical and interacting sources of memory traffic: \textit{(i)} workloads executed on RISC-V soft cores instantiated on the FPGA and \textit{(ii)} memory traffic injected from the host system via direct memory access (DMA) over the PCIe interface.
The methodology, depicted in Fig.~\ref{fig:arch_top}, allows evaluating a number of performance metrics, including memory bandwidth and latency, under diverse workloads and for different configurations of the RISC-V soft cores.
Toki advances the current state of the art through three main contributions.

\begin{enumerate}
	\item \emph{RISC-V soft cores --} Toki enables profiling the performance attainable on an FPGA by pairing HBM and a vast design space of programmable RISC-V soft cores, whose number, organization into core complexes, and architectural and microarchitectural parameters are defined at design time.
	
	\item \emph{PCIe host DMA traffic --} Beyond on-chip compute tasks, Toki supports injection of user-configurable memory traffic originating from the host system via DMA over the PCIe interface of the FPGA accelerator card, enabling realistic concurrent stress on HBM.
	
	\item \emph{Real-world application workloads --} The programmability of Toki's RISC-V compute cores makes it possible to execute any application and stress HBM with memory access patterns and workloads neglected by state-of-the-art solutions, which are instead limited to synthetic traffic generators.
\end{enumerate}

An \emph{extensive experimental campaign} targets an AMD Alveo U55C card~\cite{AMD_DS978} with a second-generation HBM (HBM2) memory,
instantiating up to 60 32-bit RISC-V Snitch~\cite{Zaruba_2021TC} compute cores and executing both microbenchmarks to
stress specific memory access patterns and a set of real-world applications from the PolyBench/C~\cite{Yuki_2014IMPACT} benchmark suite.

The whole Toki framework, encompassing all its hardware and software components and with support for widely available commercial tools and devices, is \emph{released as open source} under Apache License Version 2.0 to foster further research and ensure full reproducibility of the experiments and analyses.\footnote{Sources available at \url{https://github.com/hardware-fab/Toki}.}

\section{Toki Profiling Framework}
\label{sec:architecture}
The Toki HW/SW framework is designed for deployment on FPGA accelerator cards with a PCIe interface
that connects them to the host system, which can access HBM via DMA.
Toki's architecture can be split into three main parts, namely, the compute, memory, and DMA subsystems.

\begin{figure}[t]
	\centering
	\includegraphics[width=0.99\columnwidth]{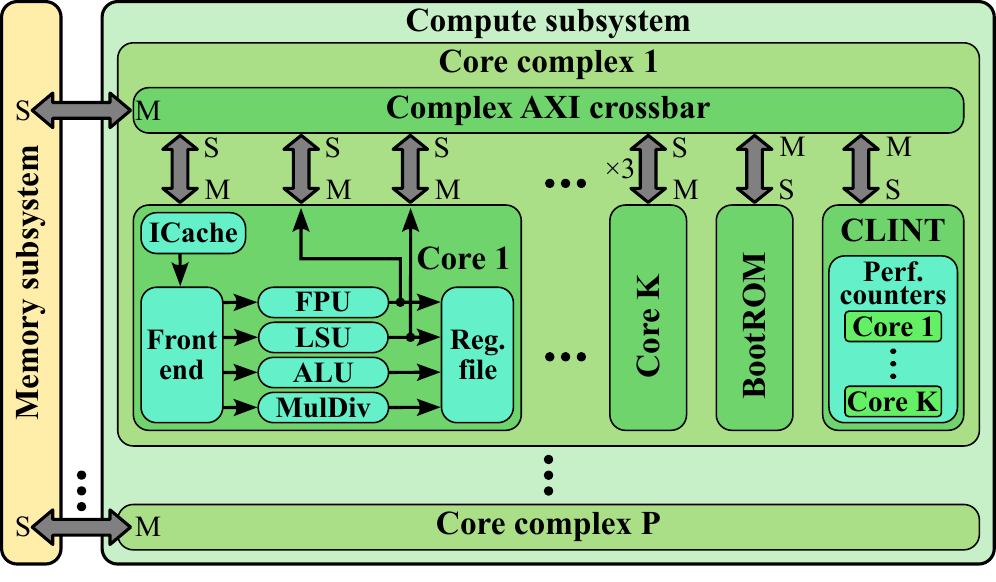}
	\caption{Detailed architecture of Toki's compute subsystem.
		\textit{Legend:} \textit{K} number of cores per core complex, \textit{P} number of core complexes, \textit{M} AXI master, \textit{S} AXI slave.}
	\label{fig:arch_compute}
\end{figure}

\subsection{Compute Subsystem}
\label{ssec:arch_compute}
The compute subsystem, shown in more detail in Fig.~\ref{fig:arch_compute}, comprises a cluster of CPU cores organized into multi-core complexes. The number of complexes (P) and the number of cores per complex (K) are configurable parameters defined at design time, offering flexibility to tailor the architecture to the user's requirements.

\begin{figure*}[t]
	\centering
	\includegraphics[width=0.94\textwidth]{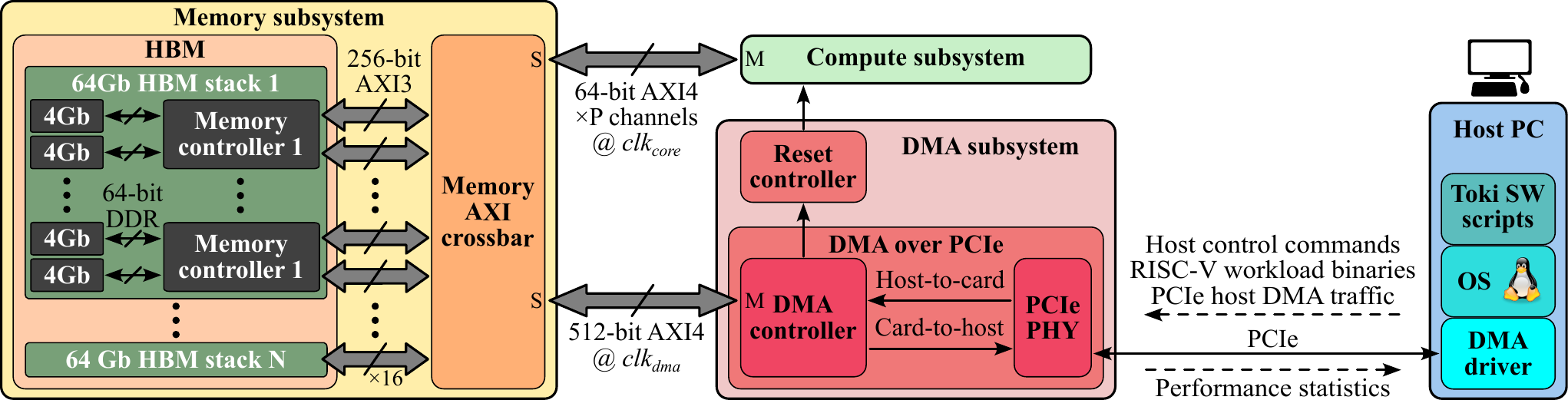}
	\caption{Detailed architecture of Toki's memory and DMA subsystems.
		\textit{Legend:} \textit{P} number of core complexes, \textit{N} number of HBM stacks, \textit{M} AXI master, \textit{S} AXI slave.}
	\label{fig:arch_memDMA}
\end{figure*}

The CPU cores of each complex are connected to an AXI crossbar that manages memory accesses. The integrated cores must expose an AXI4 master interface to communicate with the crossbar. Depending on the core design, they can feature multiple separate AXI4 channels for instruction fetch, integer load-store unit (LSU), and floating-point load-store instructions of the floating-point unit (FPU). The AXI4 interface between the cores and the crossbar operates at the core clock frequency, avoiding clock domain crossings, and is parametric in terms of data width and address width.
The compute subsystem's flexibility allows heterogeneous configurations where different core designs can be instantiated either within the same complex or across different complexes. Cores may support various cache architectures, including cacheless designs or configurations with separate or unified instruction and data caches.

Each core complex contains all the necessary support logic to operate the cores, including a BootROM and a core-local interrupt (CLINT) controller that manages timer and software interrupts.
The BootROM read-only memory stores, for each core, the boot code that initializes peripherals and jumps to the memory region assigned to the core, including both instructions and data. In addition, each core is allocated a memory region for data exchange with the host system.
The CLINT controller of each core complex includes user-configurable per-core counters dedicated
to measuring performance-related statistics such as throughput, latency, and execution time.
Both the BootROM and the CLINT controller are connected to the crossbar as AXI4 slaves with 64-bit data buses and 32-bit address buses.
The AXI crossbar in each core complex connects the compute cores to the memory subsystem and accordingly routes the memory transactions. 
It supports burst transactions, atomic operations, and up to 16 in-flight transactions per master and per slave to scale with increasing numbers of compute cores.

The compute subsystem supports the execution of any application, enabling diverse workloads and profiling scenarios.
The Toki framework includes in its open-source release a suite of user-configurable memory microbenchmarks that combine RISC-V assembly and C to implement specific access patterns, e.g., sequential, sequential with stride, and random, akin to those produced by synthetic traffic generators.
Conversely, real-world applications can also be executed to profile performance in realistic scenarios.
The compute subsystem is designed to support multi-threaded and non-uniformly parallel workloads with both structured and unstructured computational patterns, though the experimental analysis in this work focuses on applications from the PolyBench/C~\cite{Yuki_2014IMPACT} 4.2 suite.

The software stack of the Toki framework notably allows selectively disabling at run time
core complexes and cores without reprogramming the FPGA.
This enables the straightforward profiling of configurations that are smaller than
the ones physically instantiated on the FPGA.
For instance, experiments on a compute subsystem configuration with four dual-core complexes
can be equivalently carried out on a compute subsystem configuration with six quad-core complexes
by disabling two core complexes as well as two cores of each enabled core complex.

\subsection{Memory Subsystem}
\label{ssec:arch_memory}
The memory subsystem includes the HBM with memory controllers and an AXI crossbar, as shown on the left in Fig.~\ref{fig:arch_memDMA}.

The HBM consists of one or more 64Gb stacks, each containing sixteen 4Gb DRAM blocks. Each block communicates with its associated memory controller via a 64-bit pseudo channel, with all sixteen pseudo channels combining to form a 1024-bit wide HBM interface. Eight memory controllers per stack handle the translation of AXI transactions into memory transactions and issue the necessary control signals to the DRAM cells. Each controller manages two pseudo channels, corresponding to a pair of DRAM blocks.

The memory subsystem connects to the compute subsystem via the AXI crossbar, that is responsible for routing memory transactions to the appropriate memory controller. An AXI4 channel is instantiated for each multi-core complex in the compute subsystem. These channels share consistent parameter settings, including data width, address width, and burst capabilities, and operate at the core clock frequency. The crossbar interfaces with the memory controllers via two 256-bit AXI3 channels per controller.

The AXI crossbar handles protocol, data width, and clock domain conversions between the compute subsystem and the memory controllers' AXI interfaces. Moreover, it allows fine-grained control over the memory address space, enabling selective access to memory regions by different core complexes. Finally, it facilitates communication between the DMA subsystem and the HBM controllers, enabling the PCIe host to read from and write to the HBM. This functionality supports uploading compiled binaries for the compute subsystem's cores and retrieving performance statistics.

\subsection{DMA Subsystem}
\label{ssec:arch_dma}
The DMA subsystem, depicted at the center of Fig.~\ref{fig:arch_memDMA}, facilitates the communication between the host PC and the FPGA's memory subsystem. It consists of a DMA controller, a reset controller, and a PCIe physical layer (PHY) interface.

The DMA controller enables the transfer of data between the host and the FPGA's HBM, as well as the transmission of reset commands to the reset controller, which initializes the compute subsystem.
It connects to the memory subsystem through an AXI4 master interface, which is 512-bit wide and supports burst transfers and multiple in-flight transactions, and communicates with the host via the PCIe interface, connected to the FPGA's PCIe PHY layer.
The host operating system (OS) orchestrates data transfers via a dedicated DMA driver that leverages scatter-gather mechanisms to optimize large transactions, thereby avoiding the need for contiguous memory allocations.
The HBM is exposed to the host as two character devices, namely, one serving host-to-card read and write requests and the other handling card-to-host transfers.

The host system can generate memory access traffic to the HBM by issuing read and write requests with various patterns, such as random, sequential, and sequential with stride, via DMA over the PCIe interface. This interface connects the FPGA and its HBM memory to the rest of the system. The DMA subsystem enables byte granularity for read and write operations, allowing the host to interact with the HBM at runtime. Moreover, it supports loading software payloads for the compute cores and reinitializing the compute subsystem when switching workloads during profiling, with the reset controller generating accordingly a reset signal when triggered by a dedicated host command.

\section{Experimental Setup}
\label{sec:experiments}
\subsubsection*{Hardware Setup}
Our experiments targeted an AMD Alveo U55C card~\cite{AMD_DS978},
that integrates an UltraScale+ FPGA and 16GB of HBM2,
split into two 64Gb stacks each with eight memory controllers,
and that connects to the host PC via a PCIe Gen3x16 interface.

The host PC is configured with an Intel i7-10700 CPU and
64GB of DDR4-2933 memory, and runs Ubuntu 22.04.5 LTS.
The AMD Vivado ML 2023.2 toolchain was employed for
RTL synthesis and implementation, bitstream generation, and programming of the AMD Alveo U55C card.

\subsubsection*{Software Setup}
The workload executed on the CPU cores instantiated on the FPGA
included a number of real-world applications from the PolyBench/C~\cite{Yuki_2014IMPACT} 4.2 benchmark suite.
The sixteen PolyBench/C applications used in the experimental campaign
were compiled with gcc 13.2.0 from the RISC-V GNU compiler toolchain.
Workload execution is performed bare-metal, with each core of each complex
running one thread of the application under execution.

The applications' binaries are loaded at run time from the host PC through the PCIe connection
into the HBM, from which instructions are fetched by the CPU cores to be executed.
PCIe data transfers from the host system leverage AMD's XDMA driver to store the application
binaries in the HBM, retrieve performance-related statistics from the U55C card, and
stress memory with host traffic incoming through DMA over PCIe.

\subsubsection*{Toki Setup}
\begin{figure}[t]
	\centering
	\includegraphics[width=\columnwidth]{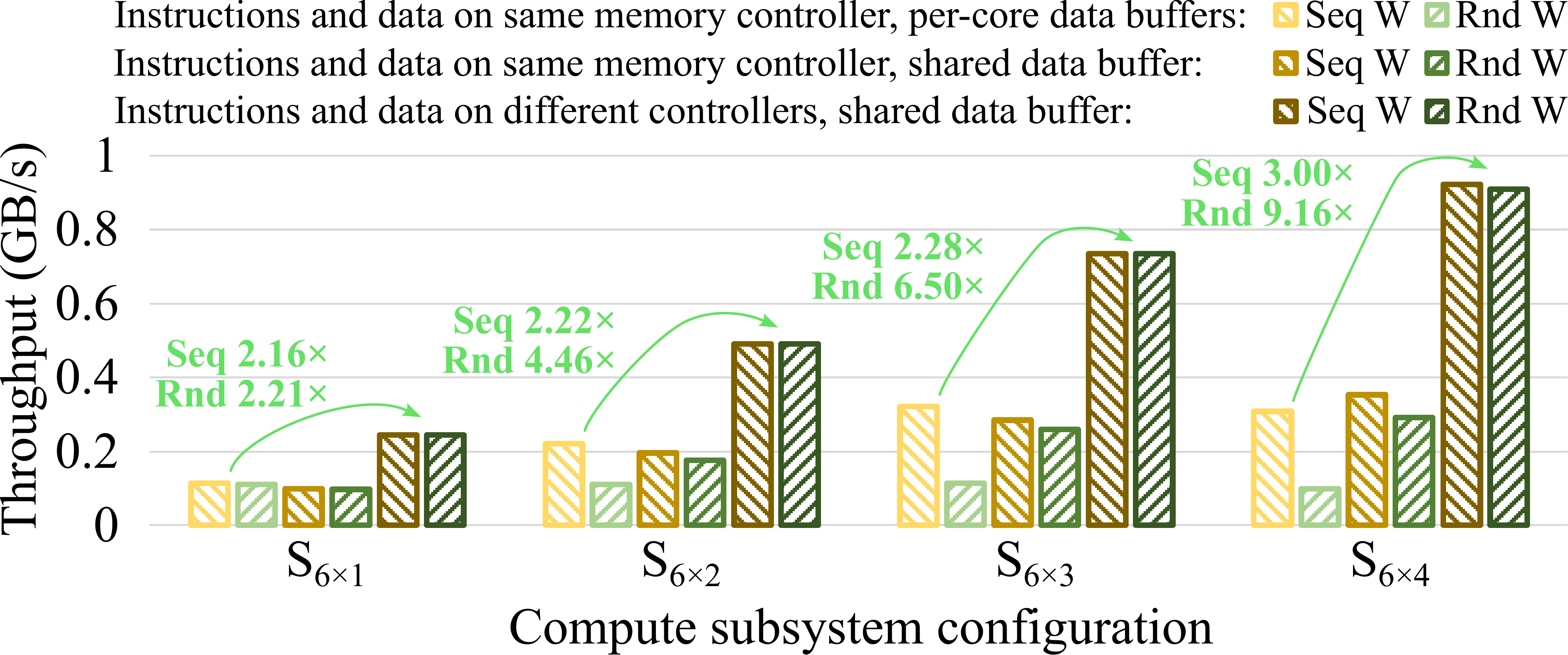}
	\caption{Memory throughput when executing sequential and random write microbenchmarks with data buffer shared among cores and with separate buffers, and with instructions and data sections mapped on same memory controller and on different ones.
	\textit{Legend:} \textit{Seq W} sequential write, \textit{Rnd W} random write.
	\textit{Note:} arrows denote throughput increase.}
	\label{fig:membench_throughput}
\end{figure}

The experimental campaign considered a variety of configurations of the Toki architecture that
differed in the number of core complexes and in the number of RISC-V cores per complex.
Snitch~\cite{Zaruba_2021TC}, a 32-bit RISC-V core that supports the RV32IMAFDC extensions
and that features the OpenHW Group's CVFPU~\cite{Mach_2021TVLSI},
is used as the compute core, configured with a 4kB instruction cache and no data cache.
Synthesis and implementation targeted operating frequencies of
50MHz~(\textit{clk\textsubscript{core}} in Fig.~\ref{fig:membench_throughput}) for the compute subsystem
and 250MHz~(\textit{clk\textsubscript{dma}}) for the memory and DMA subsystems.
The Toki architecture also makes use of AMD's XDMA~\cite{AMD_PG195} and HBM~\cite{AMD_PG276}
IPs as its DMA and memory controllers.

Toki supports instantiating on the Alveo U55C card up to 60 Snitch cores organized into up to twelve clusters, providing a maximum memory throughput of 4.8GB/s and with a maximum resource utilization of \num{1184903} lookup tables (LUTs), \num{875554} flip-flops (FFs), \num{900} DSPs, \num{820} 36kB blocks of block RAM~(BRAM), and \num{5} 288kB blocks of UltraRAM~(URAM).
As an alternative compute core, the application-class
64-bit RISC-V CVA6~\cite{Zaruba_2019TVLSI} core features both configurable instruction and data caches
and operates at 100MHz.

\begin{table}[t]
	\centering
	\setlength{\tabcolsep}{2pt}
	\caption{Resource utilization breakdown of Toki architecture.}
	\label{tab:exp_areaBreakdown}
	{
		\fontsize{9pt}{9pt}\selectfont
		\begin{tabular}{lrrrrr}
			\toprule
			\textbf{Design component}                        & \textbf{kLUT} & \textbf{kFF} & \textbf{DSP} & \textbf{BRAM} & \textbf{URAM} \\ \midrule
			\textbf{Compute subsystem}                       &         382.0 &        252.4 &          360 &           288 &             0 \\
			$\rightarrow$ Core complex (4\texttimes )        &          95.5 &         63.1 &           90 &            72 &             0 \\
			\ \ \ \ $\rightarrow$ Snitch core (6\texttimes ) &          13.5 &          6.3 &           15 &            12 &             0 \\
			\ \ \ \ $\rightarrow$ BootROM                    &           0.3 &          0.1 &            0 &             0 &             0 \\
			\ \ \ \ $\rightarrow$ CLINT controller           &           0.2 &          0.2 &            0 &             0 &             0 \\
			\ \ \ \ $\rightarrow$ Complex AXI crossbar       &          14.0 &         25.0 &            0 &             0 &             0 \\ \midrule
			\textbf{Memory subsystem}                        &          16.0 &         29.0 &            0 &             6 &             0 \\
			$\rightarrow$ Memory controllers                 &           1.5 &          1.6 &            0 &             6 &             0 \\
			$\rightarrow$ Memory AXI crossbar                &          14.5 &         27.4 &            0 &             0 &             0 \\ \midrule
			\textbf{DMA subsystem}                           &          59.6 &         66.0 &            0 &           112 &             5 \\
			$\rightarrow$ DMA over PCIe                      &          59.5 &         65.9 &            0 &           112 &             5 \\
			$\rightarrow$ Reset controller                   &           0.1 &          0.1 &            0 &             0 &             0 \\ \midrule
			\textbf{Total}                                   &         457.6 &        347.4 &          360 &           406 &             5 \\ \bottomrule
		\end{tabular}
	}
	\begin{minipage}{\linewidth}
		\vspace{1ex}
\footnotesize \textit{Note:} Toki architecture
deployed on AMD Alveo U55C card with compute subsystem in S\textsubscript{6\texttimes4} configuration,
i.e., 24 Snitch cores organized into four core complexes.
	\end{minipage}
\end{table}

\begin{figure*}[t]
	\centering
	\includegraphics[width=\textwidth]{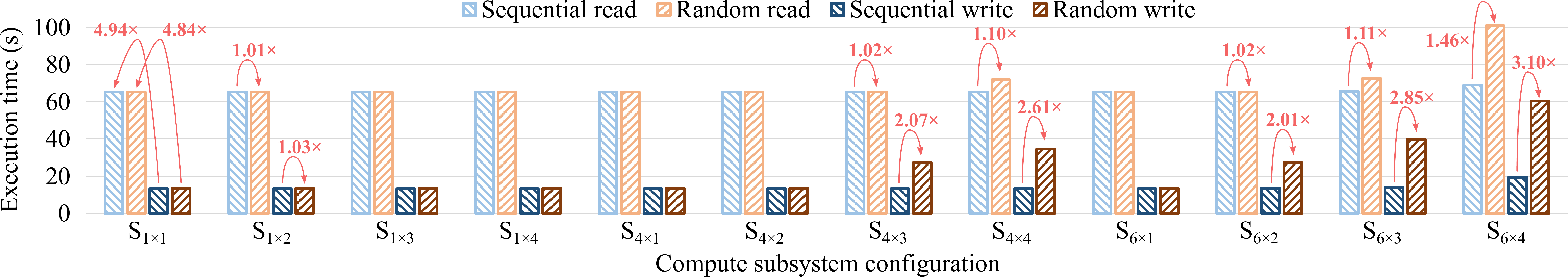}
	\caption{Execution time, expressed in seconds (s), on each compute core of memory microbenchmarks running on S\textsubscript{\textit{K}\texttimes \textit{P}} Toki configurations with \textit{K} Snitch cores per complex and \textit{P} core complexes.
	\textit{Note:} arrows denote performance degradation.}
	\label{fig:membench_execTime}
\end{figure*}

\section{Experimental Results}
\label{sec:results}
Without loss of generality, we showcase Toki's effectiveness with an experimental analysis that profiles the performance achievable in a worst-case scenario, where all memory accesses stress solely one memory controller of one HBM stack of the U55C card, for instance, in Fig.~\ref{fig:arch_memDMA}, Memory controller 1 of the 64Gb HBM stack~1.
The experiments execute memory microbenchmarks and real-world applications on the compute cores and include the injection of memory traffic from the PCIe host.
We refer to Toki instances with \textit{K} Snitch cores per complex
and \textit{P} core complexes as S\textsubscript{\textit{K}\texttimes\textit{P}}.
For instance, the S\textsubscript{6\texttimes4} configuration features
four core complexes each of whom includes six Snitch cores.
Table~\ref{tab:exp_areaBreakdown} lists the post-implementation resource utilization of
Toki in its S\textsubscript{6\texttimes4} configuration, breaking it down
among the three subsystems and their components.

\subsubsection*{Memory Microbenchmarks -- Execution Time}
The experimental campaign makes use of four of the memory microbenchmarks included in Toki,
stressing HBM from the compute cores with sequential and random reads
and sequential and random writes.
The microbenchmarks involve reading and writing 256MB of data, executing 256 iterations
of read and write operations on a 1MB data buffer.
All memory sections dedicated to instructions and data of the application under execution are stored in HBM blocks associated to the same memory controller and each core has its own data buffer.

Fig.~\ref{fig:membench_execTime} depicts the execution time on each core of
the four microbenchmarks running on a set of Toki configurations with
different numbers of core complexes and cores per complex.
Such metric tends in general to increase with the number of compute cores, as the latter contend access to the shared memory.
The per-core execution time is almost equivalent among sequential and random reads up to twelve total compute cores, while sequential and random writes perform similarly up to eight.
Conversely, once such thresholds are exceeded, the gap in performance between sequential and random accesses starts rapidly increasing, with the random accesses microbenchmark being up to 3.10\texttimes\, slower than the sequential one in the write use case on the S\textsubscript{6\texttimes4} configuration. 
Write operations are notably up to 4.94\texttimes\, faster than read ones, with the largest gap achieved with the lowest core count, i.e., S\textsubscript{1\texttimes1}.

\subsubsection*{Memory Microbenchmarks -- Throughput}
We evaluate then how memory throughput varies by sharing first the 1MB data buffer among all the compute cores to increase the data locality, and then by moving the data buffer to memory blocks managed by a different memory controller to remove the contention between instruction and data accesses.
The experiments in Fig.~\ref{fig:membench_throughput} focus, without loss of generality, on configurations with six Snitch cores per complex and a number between one and four of complexes stressed only with writes.
The experiments compare the total memory throughput, expressed in GB/s, achieved first by executing the sequential and random write microbenchmarks in the baseline conditions as in Fig.~\ref{fig:membench_execTime}, then sharing the 1MB data buffer among all the cores, and finally with separate memory controllers for data and instructions.

Notably, in the baseline scenario with instructions and data on the same memory controller and with per-core 1MB data buffers, the single memory controller is unable to provide a larger throughput as the core count increases with the least favorable memory access pattern, i.e., random writes. Memory throughput even slightly decreases from S\textsubscript{6\texttimes1}'s 0.11GB/s down to S\textsubscript{6\texttimes4}'s 0.10GB/s due to the controller's saturation coupled with the increasing contention.

Sharing the data buffer among the cores is shown to raise the throughput with a more meaningful improvement at higher core counts and with random accesses, e.g., the S\textsubscript{6\texttimes4} configuration's random write throughput is 2.93\texttimes\, higher with a shared data buffer than with per-core ones.
Such modification also reduces the gap between sequential and random accesses at higher core counts.

\begin{figure*}[t]
	\centering
	\includegraphics[width=\textwidth]{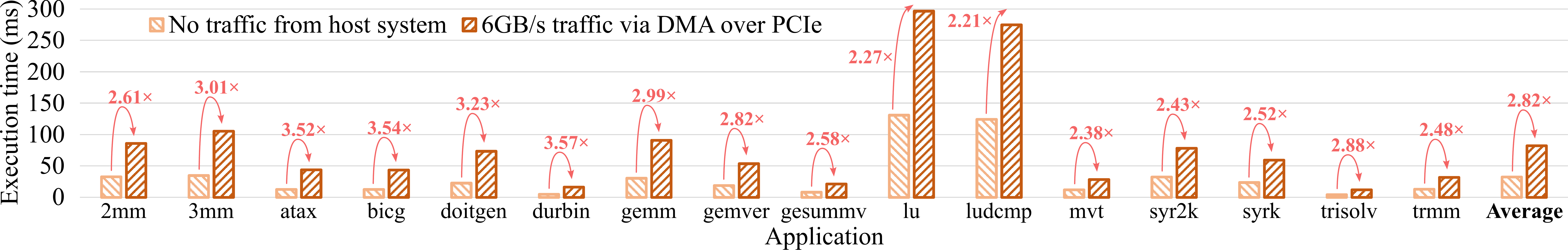}
	\caption{Execution time, expressed in milliseconds (ms), of each PolyBench/C application in S\textsubscript{1\texttimes1} configuration without and with 6GB/s memory traffic from PCIe host system.
	\textit{Note:} arrows denote performance degradation due to PCIe host traffic.}
	\label{fig:polybench_perApp}
\end{figure*}

\begin{figure*}[t]
	\centering
	\includegraphics[width=\textwidth]{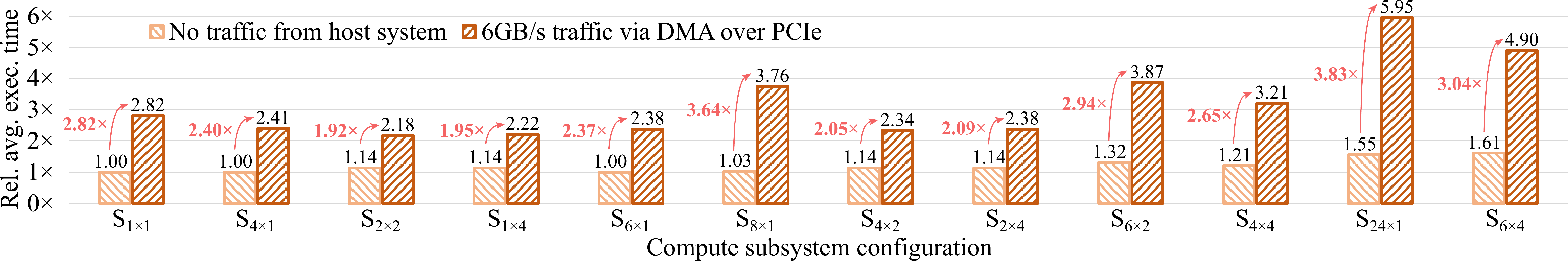}
	\caption{Relative average time for execution on each compute core of PolyBench/C applications without and with memory traffic from PCIe host system. \textit{Note:} normalized against S\textsubscript{1\texttimes1} no-traffic scenario~(Fig.~\ref{fig:polybench_perApp}), arrows denote performance degradation.}
	\label{fig:polybench_compare}
\end{figure*}

Separating instructions and data sections on different memory controllers further increases the memory throughput, with a higher improvement at higher core counts and with random accesses. For instance, such modification increases the S\textsubscript{6\texttimes4} configuration's throughput by 2.61\texttimes\, and 3.13\texttimes\, for sequential and random writes, respectively, with a total improvement on the baseline scenario of 3.00\texttimes\, and 9.16\texttimes.
In this final scenario, scaling from one to four the number of six-core complexes produces an almost linear increase in the total memory throughput, with the S\textsubscript{6\texttimes2}, S\textsubscript{6\texttimes3}, and S\textsubscript{6\texttimes4} configurations having throughput values 2.0\texttimes, 3.0\texttimes, and more than 3.8\texttimes\, higher than the S\textsubscript{6\texttimes1} one for both sequential and random writes.
The improvement is indeed equal to the maximum theoretical one with two and three clusters and close to it with four.

\subsubsection*{Real-World Applications -- Execution Time}
Fig.~\ref{fig:polybench_perApp} depicts the execution time, on the Toki S\textsubscript{1\texttimes1} configuration, of each of the sixteen considered PolyBench/C applications when there is no traffic from the host system and conversely when the host system generates and sends to HBM a write-only traffic of 6GB/s via DMA over PCIe, contending the shared memory resource with the compute subsystem.
When executing on one Snitch core, the execution of PolyBench/C applications takes in the order of tens to hundreds of milliseconds.
When 6GB/s traffic is incoming from the host, their execution takes 2.82\texttimes\ on average and up to 3.57\texttimes\ longer than in the no-traffic scenario, with the largest slowdown occurring for the \textit{durbin}, \textit{bicg}, and \textit{atax} applications.

\subsubsection*{Real-World Applications -- PCIe Host Traffic}
Fig.~\ref{fig:polybench_compare} depicts the relative average time for the execution on each compute core of the sixteen applications on a wide range of Toki instances, compared to the no-traffic scenario on the S\textsubscript{1\texttimes1} configuration which acts as the baseline and whose bar has a value of 1\texttimes\, in the chart.

The injection of traffic from the host system is shown to drastically slow down the execution on the compute cores, also due to the higher throughput of 6GB/s via DMA over PCIe compared to the lower one of compute cores even when many of them are instantiated, e.g., the total throughput of compute cores is lower than 1GB/s with both optimizations applied as shown in Fig.~\ref{fig:membench_throughput}.

In particular, the impact of PCIe host traffic ranges from a 1.92\texttimes\, increase in the average execution time of PolyBench/C applications up to a 3.83\texttimes\, slowdown. The latter value corresponds to a S\textsubscript{24\texttimes1} compute subsystem, i.e., the configuration considered in the experimental evaluation with the highest total core count and number of cores per complex, a hierarchical organization of the compute cores that produces the highest contention of shared resources.

\section{Related Work}
\label{sec:related}
\begin{table}[t]
	\centering
	\footnotesize
	\setlength{\tabcolsep}{2pt}
	\caption{Feature comparison of HBM profiling frameworks.}
	\label{tab:relatedwork_comparison}
	\begin{tabular}{lccccc}
		\toprule
		                                  &                     &       \textbf{Memory}       &     \textbf{HBM}     &     \textbf{DRAM}     &               \\
		\textbf{Feature}                  &   \textbf{Shuhai}   &      \textbf{Sandbox}       &   \textbf{Connect}   &    \textbf{Bender}    & \textbf{Toki} \\ \hline
		Synthetic traffic generation      &      \fullcirc      &          \fullcirc          &      \fullcirc       &       \fullcirc       &   \fullcirc   \\
		Multiple pseudo-channel support   &      \halfcirc      &          \fullcirc          &      \fullcirc       &       \halfcirc       &   \fullcirc   \\
		Random access patterns            &     \emptycirc      &          \fullcirc          &      \fullcirc       &       \fullcirc       &   \fullcirc   \\
		Programmable multi-core compute   &     \emptycirc      &         \emptycirc          &      \emptycirc      &       \fullcirc       &   \fullcirc   \\
		RISC-V open-source cores          &     \emptycirc      &         \emptycirc          &      \emptycirc      &      \emptycirc       &   \fullcirc   \\
		Real-world application workloads  &     \emptycirc      &         \emptycirc          &      \emptycirc      &      \emptycirc       &   \fullcirc   \\
		Host PCIe/DMA traffic injection   &     \emptycirc      &         \emptycirc          &      \emptycirc      &      \emptycirc       &   \fullcirc   \\
		Concurrent on-chip, host accesses &     \emptycirc      &         \emptycirc          &      \emptycirc      &      \emptycirc       &   \fullcirc   \\ \bottomrule
	\end{tabular}
	\begin{minipage}{\linewidth}
		\vspace{1ex}
		\footnotesize \textit{Legend:} \fullcirc~fully supported, \halfcirc~partially supported, \emptycirc~not supported.\\
		\textit{Note:} Toki and DRAM Bender generate synthetic traffic through programmable cores, allowing reconfigurable access patterns distinct from fixed hardware generators.
	\end{minipage}
\end{table}

The existing solutions for the evaluation of HBM
performance, compared with Toki in Table~\ref{tab:relatedwork_comparison},
consider synthetic scenarios with traffic generated
according to some configurable memory access patterns and do
not target real-world applications executing on multi- or many-core
CPUs, including RISC-V-based ones.
Moreover, they treat FPGAs with HBM as standalone devices rather
than as part of more complex heterogeneous systems, ignoring
for instance the impact given by concurrent accesses to HBM
from logic instantiated on the FPGA and from the host system.

Shuhai~\cite{Huang_2022TC} and Memory Sandbox~\cite{Perdomo_2024SBAC-PAD} both use synthetic traffic generators to benchmark
HBM2 performance on AMD’s Alveo U280 card.
Shuhai~\cite{Huang_2022TC} focuses on a single HBM pseudo-channel with fixed-stride sequential read-only or write-only accesses,
failing to consider coherent data traffic, random access patterns, memory accesses from the host system, and realistic workloads.
Memory Sandbox~\cite{Perdomo_2024SBAC-PAD} stresses multiple pseudo-channels, supports simultaneous accesses to the same memory region, and enables both sequential and random access patterns.
HBM Connect~\cite{Choi_2021FPGA} implements both the traffic generators and the interconnect using HLS, achieving a drastically lower bandwidth, and
\cite{Lu_2021FPGA} also focuses on HLS-based accelerators deployed on FPGA platforms with HBM.
DRAM Bender~\cite{Olgun_2023TCAD} employs a custom-ISA core to execute fine-grained memory tests, enabling precise control for low-level characterization analyses of HBM2 such as Rowhammer~\cite{Mutlu_2019TCAD} vulnerability. While effective for standalone testing and isolated DRAM analysis, it overlooks performance characterization in broader system contexts.

\section{Conclusions}
\label{sec:conclusions}
This paper introduced Toki, the first HW/SW framework designed to profile HBM performance in FPGA-based systems with RISC-V soft cores and PCIe traffic incoming from the host via DMA.
Toki's design is intentionally modular and extensible. The compute subsystem 
can accommodate diverse processor designs beyond Snitch and CVA6,
including many-core overlay processors and soft GPGPUs.
Similarly, the framework is portable across FPGA platforms with HBM support,
with current validation on the AMD Alveo U55C and planned support
for future generations and alternative devices.

Looking forward, Toki provides a foundation for systematic evaluation of emerging 
FPGA architectures, memory technologies, and processor designs. By releasing the 
whole framework as open source, we invite the community to extend and adapt Toki to their 
specific research needs and contribute new insights to the field.

\begin{acks}

\includegraphics[height=1em]{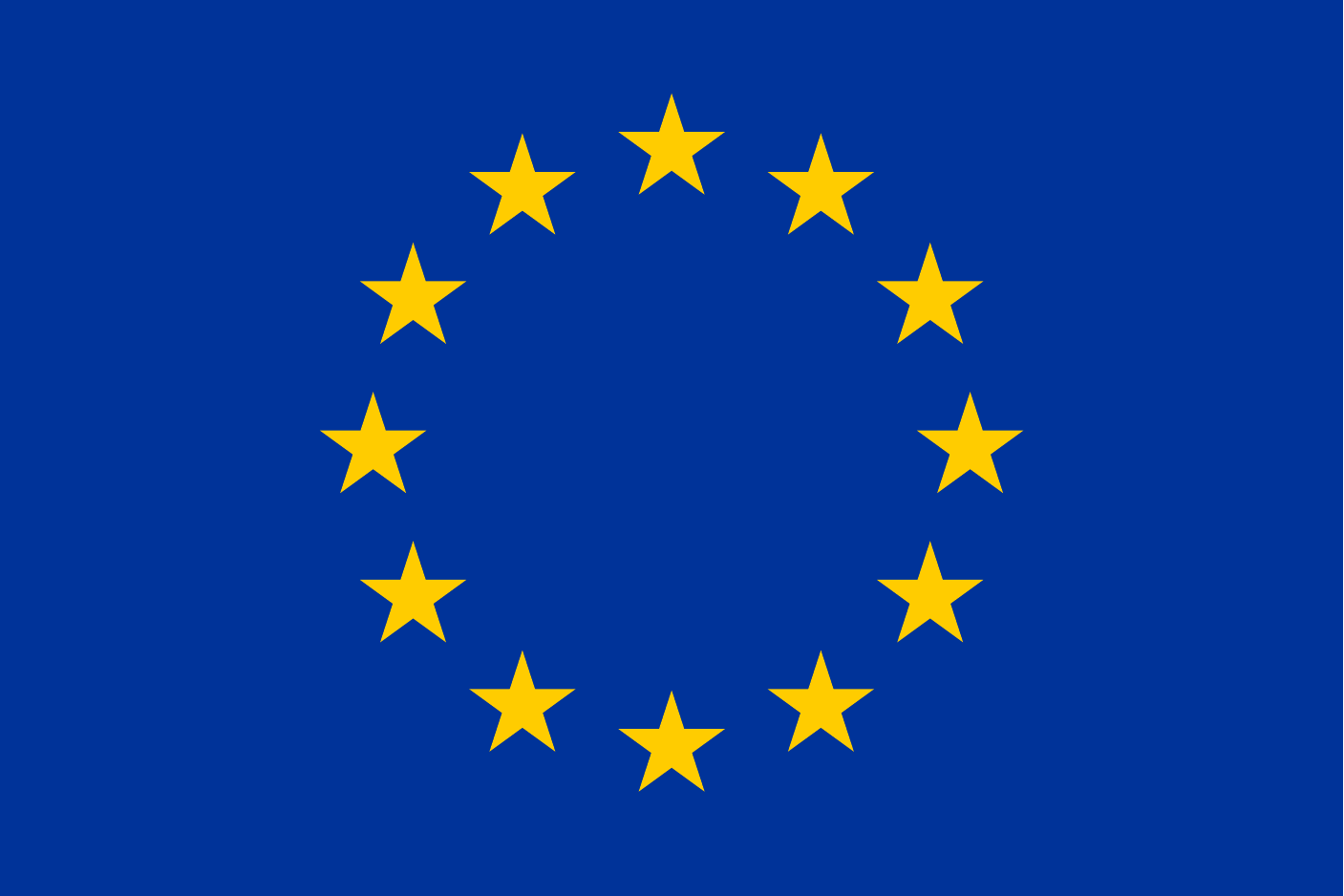}
Funded by the European Union. Views and opinions expressed
are however those of the author(s) only and do not necessarily
reflect those of the European Union or HaDEA. Neither
the European Union nor the granting authority can be held responsible for them.
The authors thank the AMD University Program for providing
the AMD Alveo U55C card used in this work.
\end{acks}

\bibliographystyle{ACM-Reference-Format}
\bibliography{refs}
	
\end{document}